# Coupling-controlled Dual ITO Layer Electro-Optic Modulator in Silicon Photonics

MOHAMMAD H. TAHERSIMA[1*], ZHIZHEN MA[1*], YALIANG GUI[1], MARIO MISCUGLIO[1], SHUAI SUN[1], RUBAB AMIN[1], HAMED DALIR[2], VOLKER J. SORGER[1]

[1]George Washington University, 800 22nd Street NW, Washington, DC 20052, USA
[2]Omega Optics, Inc. 8500 Shoal Creek Blvd., Bldg. 4, Suite 200, Austin, Texas 78757, USA

*Corresponding author: sorger@gwu.edu

**Abstract**

*Electro-optic signal modulation provides a key functionality in modern technology and information networks. Photonic integration has enabled not only miniaturizing photonic components, but also provided performance improvements due to co-design addressing both electrical and optical device rules. However the millimeter-to-centimeter large footprint of many foundry-ready photonic electro-optic modulators significantly limits scaling density. Furthermore, modulators bear a fundamental a frequency-response to energy-sensitive trade-off, a limitation that can be overcome with coupling-based modulators where the temporal response speed is decoupled from the optical cavity photo lifetime. Thus, the coupling effect to the resonator is modulated rather then tuning the index of the resonator itself. However, the weak electro-optic response of silicon limits such coupling modulator performance, since the micrometer-short overlap region of the waveguide-bus and a microring resonator is insufficient to induce signal modulation. To address these limitations, here we demonstrate a coupling-controlled electro-optic modulator by heterogeneously integrating a dual-gated indium-tin-oxide (ITO) phase shifter placed at the silicon microring-bus coupler region. Our experimental modulator shows about 4 dB extinction ratio on resonance, and a about 1.5 dB off resonance with a low insertion loss of 0.15 dB for a just 4 μm short device demonstrating a compact high 10:1 modulation-to-loss ratio. In conclusion we demonstrate a coupling modulator using strongly index-changeable materials enabling compact and high-performing modulators using semiconductor foundry-near materials.*

**INTRODUCTION**

Integrated electro-optic (EO) modulators perform key applications in telecom and data communications [1], future on-chip and inter-chip photonic interconnects for multicore microprocessors and memory systems [2,3], RF and analog signal processing such as photonic A/D conversion [4] and in sensors [5]. Monolithic integration, mainly in silicon photonics, enabled i) densifying photonic networks compared to discreetly-packaged components, ii) reduced device power-consumption [6], and iii) enabled a platform approach for cost- and density-scaling due to synergies arising largely processing Silicon-based components. The weak EO properties of silicon [7-9], however, result in order of millimeter-to-centimeter large footprints, and thus further density scaling, which was a major driver for the chip industry for decades [10]. The performance metrics for modulators [11-13] are high speed [5, 14,15], high sensitivity/energy efficiency [6, 12,13,17-19], high extinction and compact footprint [15-19].

Higher modulator performance can be achieved by increasing the light-matter-interaction [11-13]; this can for instance be achieved by resonances (microcavity, microring) allowing for increased sensitivity (energy efficiency) due to multiple round trips of light the cavity on resonance, and have demonstrated Gbps rates [20,21]. Alternatively the higher group indices could be used such as in plasmonics [15,16]. Another strategy is to rely on strongly index-changing materials beyond silicon such as graphene [11,14, 18, 22] or transparent conductive oxides [16, 19, 23], or combinations of resonant effects with emerging materials [11, 24]. Heterogeneous integration [7-9] has been introduced as a possible route to continue device scaling [25] without loosing modulator signaling performance; here the idea is to utilize the foundry established silicon (or III-V) platforms for passive waveguide parts, but to not use the same material for active light-manipulation. In this regard, plasmonics (and photonic-plasmon hybrids) can provide ultimate scaling, but the tradeoff is loss; that is, even the light-matter-enhanced optical mode and higher group index in plasmonics that remarkably enables functional submicron-small devices comes at a cost of (usually) about -5 dB insertion loss. Here we argue, that such loss is not acceptable with scale-up circuit where component cascadability is key, thus plasmonics is ruled out. In contrast, we follow an approach targeting a design-regime that lies between photonics (low loss & low ER) and plasmonics (high ER, high loss), namely hetero-generous integration offering a sweet spot of a) decent signal quality < -10 dB and b) low insertion loss ~ - 1dB. Here we select Indium-tin-oxide (ITO) as the EO material and integrate an ITO-oxide-ITO capacitive gate-stack atop a silicon photonic waveguide to form an all-photonic optical mode. The choice for ITO was threefold; i) ITO belongs to the class of transparent conductive oxides (TCO) and is currently used massively by both the high tech and solar-cell industry for resistive touch screens such as in smart phones, and transparent low-resistive front-end contact in photovoltaics, respectively [26-28]. Hence it might enter foundry processes sooner (especially in lower TLR foundry's such as AIM photonics) than other exotic materials such as 2-dimentional materials. ii) While process control of ITO has been challenging due to intrinsic material complexities, we have yet recently demonstrated a holistic (electronic and optical) approach to both precisely and repeatedly control ITO's parameters [29]. iii) the carrier concentration modulation can be in the order of $\sim 10^{20}$ cm$^{-3}$, which is about 100x enhanced compared to silicon [14,19].

However, there is a well-known tradeoff in sensitivity vs. modulation bandwidth in resonant modulators [30,31]; a higher quality (Q)-factor resonator enables a lower modulation energy due to a narrow spectral linewidth, but the corresponding longer cavity lifetime limits modulation response to low frequencies. This is a fundamental limitation and is valid for all conventional resonant modulator designs where the cavity resonance (or loss) are modulated. However, this limitation is broken when the coupling strength between the waveguide bus to the resonator is modulated instead of the resonance frequency of the resonator itself [32-37].

Here we follow the combined approach of utilizing a) heterogeneous integration if ITO-based for phase shifting, b) the concept of coupling modulation, c) in silicon photonics. There are a few novelties in this work, worth highlighting briefly upfront; i) we introduce a first ITO-oxide-ITO push-pull modulation, ii) use ITO away from the ENZ point to reduce losses, iii) demonstrate the first TCO-based coupling modulator. We find an efficient performance of the device of 4 dB on resonance, and a 10:1 ER/IL ratio off-resonance, which includes a low 0.15 dB insertion loss of this 4 micrometer short phase shifter.

Based on the location of the active (EO altered) region on a ring resonator there are two device types: a) interactivity device b) coupling device (Fig. 1). Each modulation mechanism determines which transmission response parameter is altered; (a) index modulation: phase $\phi$ varies in time but self-coupling factor $r$ and cross-coupling factor $k$ are constant, and (b) coupling modulation: where $r$ and $k$ vary, but $\phi$ is constant. In general phase modulators are rather effective in achieving the high extinction ratio (ER) since their resonance sharpness is almost unaffected by altering the real part of the refractive index; thus, ER can be as large as the difference between the maximum transmission and the dip of transmission resonance, resulting in spectral sharp 'peaks' (Fig. 1a, c). Although micro ring resonator (MRR)-based index-shift modulators offer large modulation depth at a relatively small footprint cost (in the order of 100's of um$^2$), they show narrow spectral bandwidth and are sensitive to temperature and process variations (Please see Supplementary materials). Loss modulation on the other hand, has a larger spectral bandwidth but it also suffers from a higher insertion loss. In contrast, coupling modulation can have relatively broadband modulation response. In coupling modulation, the active modulation region can be significantly smaller in dimension due to sensitivity of the coupling region to changes in the refractive index of the coupling gap (Fig. 1b, d). This leads to a reduced capacitance while preserving ER.

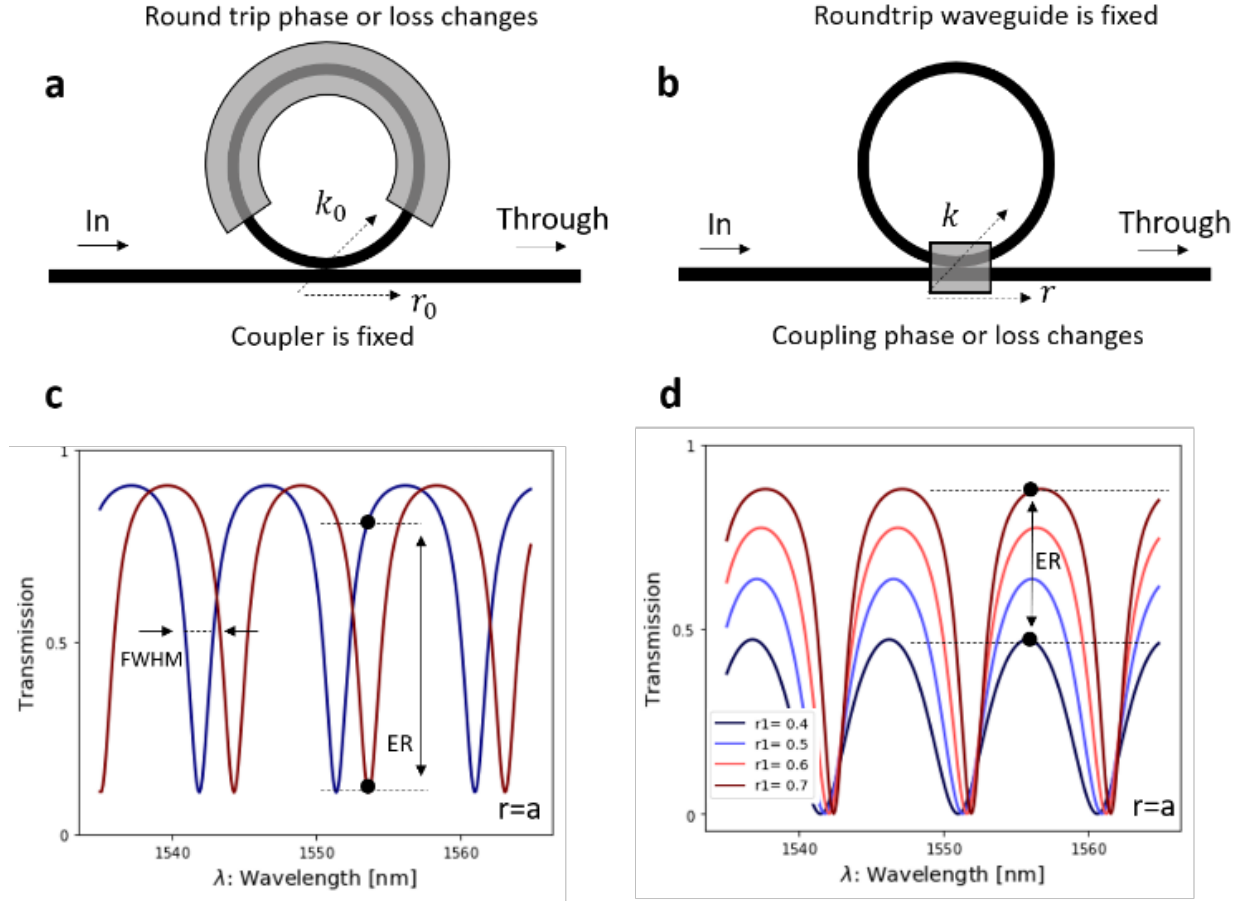

*Figure 1*: Comparison of two theoretical type of modulation mechanisms for ITO ring modulator. Schematic layout of interactivity ring resonator based electro-optic devices (a) and coupling based electro-optic devices (b). In reservoir coupling configuration, the change in effective index of coupling mode changes the coupling factor between the waveguide bus and ring resonator to modulate the optical transmission in the through port. Spectral response of transmission in the through port for an interactivity modulator (c) and a coupling region cladded ring resonator (d). When electrostatic charge is induced on the electro-optic material to change the coupling factor between bus and ring, leading to a broad band modulation effect of the through signal. In addition, there will be a small shift in resonant wavelength due to the small difference between the device coverage on the ring compared to the device coverage on the bus waveguide (effective phase shift length $\Delta L_{eff}$).

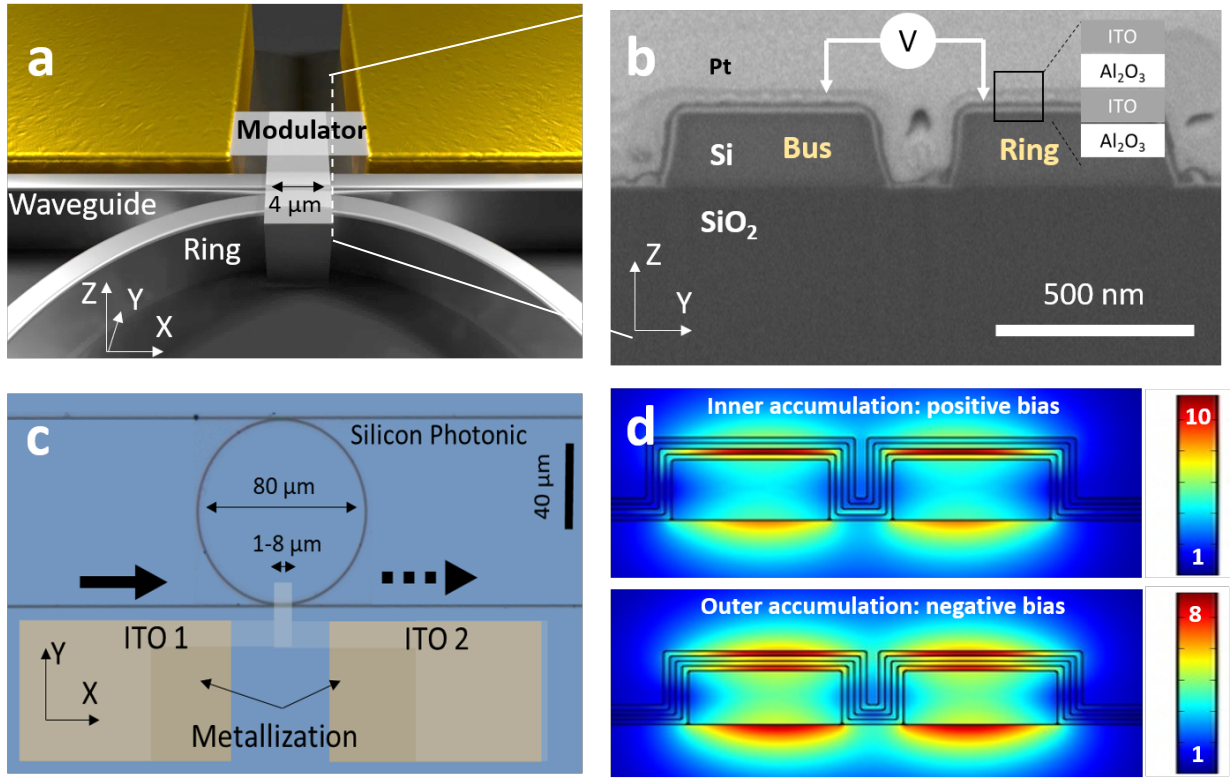

*Figure 2.* a) schematic of wavelength scale dual ITO capacitor coupling modulator. The device length is 4 μm long, b) FIBSEM cross-section illustrating the device layers and operating mechanism of the active region. An optical beam is modulated by modifying the coupling factor between two neighboring waveguides via applying an electrical bias across the double ITO layer capacitor. In the coupling region, two waveguides are spaced by a gap distance g = 20 nm, covered by an active device region. In the active coupling region, we bias two ~ 20 nm thin ITO layers separated by 10 nm of $Al_2O_3$ in a capacitive push-pull configuration to modulate the photonic (non-plasmonic) coupler region. c) Fabricated rings have a radius $r$ = 40 μm and a single pass length $L_c$ = 251.2 μm. The complete electro-optic device. Each ITO layer is extended to be connected to a 50 nm Ti/Au contact pad to carry electrical signal to the modulator. Also, each optical waveguide is connected to a grating coupler to send and receive optical signals. d) corresponding TM mode-field profile for positively and negatively biased dual ITO modulator calculated by two-dimensional Mode simulations, in which eigenmode analysis was performed on the cross-section of the device for calculating the effective mode index. Under the inner accumulation biasing condition, the mode overlap with the inner ITO layer is maximized ($\Gamma$ = 0.01) resulting in higher modulation compared to an outer accumulation condition.

Our dual-gated ITO modulator consists of a classical silicon MRR-bus configuration, but the ITO-gated phase shifter is placed at the coupling region instead on the ring only (Fig. 2a-c). The measured capacitance is 161 fF with an ITO film resistance of few tens of Ohm (see methods section), which can potentially enable beyond modulation speeds up to and possibly even beyond hundreds of GHz, with an energy per bit of the order of 100's of fJ/bit. The device material stack results in a 50 nm thick cladding on the coupling region, which effects the propagating TM mode waveguide. In detail, the coupling region is capped with the electro-optic material, at which an electrical bias is applied and according to the bias polarity, carriers are either depleted or accumulated (Fig 2 d), and consequently altering the effective index of the propagating mode in both waveguides, as well as the gap between the two. This allows index tuning of the coupling factor of the ring cavity by applying a voltage. The induced change in the carrier concentration of ITO layers by applying electrical bias tunes the refractive index of ITO layers, and consecutively the coupling factor, $\kappa$, between two bus waveguides.

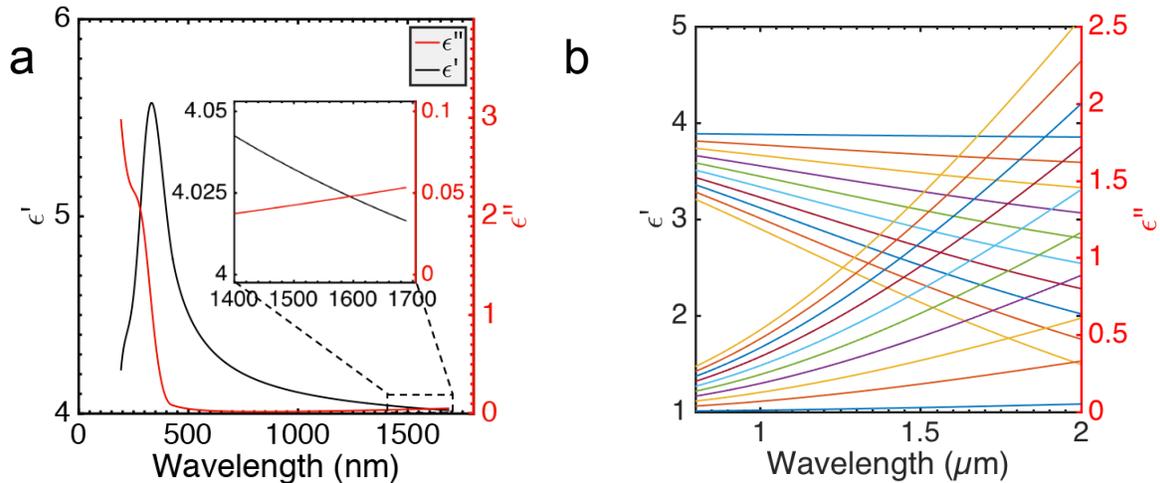

*Figure 3.* Experimental verified ITO permittivity data [29] a) Measured complex permittivity spectral response of the ITO film measured by spectroscopic ellipsometry. The ITO layer is deposited by RF sputtering, using a 1:1 (50 sccm) Oxygen to Argon flow-rate ratio for 1000s. b) Drude Model of the ITO permittivity actively modulated by electrostatic doping from $1 \times 10^{19}$ up to $5 \times 10^{20}$ $cm^{-3}$

In order to understand the observed modulation performance, we performed self-consistent experimental verifications in a holistic manner see ref [29] for details. In brief, we repeated 4-probe, Hall-bar, profilometry and ellipsometry and only collected those data-points that showed a consistency across all methods. We find a rather low imaginary part of the permittivity and consequently small extinction coefficient in the near IR region. The reason lies in the absence of a post-deposition thermal treatment, which simultaneously restores the crystallinity of the film and activates charge carriers inside ITO, reducing the overall film resistivity. In the IR region, the spectral response of the ITO film is dominated by intra-band transitions (Drude Model), although a low carrier concentration a is responsible for the rather flat refractive index response. The initial carrier concentration (without any active gating) is found to be $1.5 \times 10^{18}$ cm$^{-3}$, with a high scattering time of approximately 1 fs. Using the spectroscopic ellipsometry fitting parameter, the modulated carrier concentration, modeled by Drude model (Fig. 3b) shows a 0.2 variation of the real part of the permittivity for a high carrier concentration of $2.8 \times 10^{20}$ cm$^{-3}$, which is significantly higher but achievable in similar studies [19]

To illustrate the modulation performance of this coupling modulator, the effective refractive index for the photonic mode is altered by applying a bias between the two ITO layers (Fig.4). Extinction ratio plot are plotted by $ER = 10 \log \left( \frac{T_V}{T_{V0}} \right)$, where $T_V$ and $T_{V0}$ are output optical transmission for biased and unbiased devices. We experimentally show up to ER = 4 dB on resonance and 0.15 dB insertion loss (IL) over the wide bandwidth region of ring resonator spectral response. On resonance, the extinction ratio reaches 4 dB (ER/IL = 26) for the 4 µm device at 1552 nm wavelength due to the shift in resonance frequency of the ring as we apply a bias to ITO layers. The low insertion loss can be understood by low effective extinction coefficient of the propagating mode since the active material is a low loss material state (Fig. 3). The measured transmittance over the wavelength of five representative devices that differ in device length ($L$ = 1-16 µm) (supplementary information) highlights that the device length does not affect the quality factor of the ring (~1400). This is expected because the deposited ITO material does not have a significant extinction coefficient at operating wavelength ($\varepsilon''$(1550 nm) < 0.05). However, changing the length of the active device results in shift in resonance frequency $\Delta\phi$ following $\Delta\phi = \frac{2\pi}{\lambda}\Delta n_{eff}\Delta L$, where $\Delta L$ is the active length of phase shifter and $\Delta n_{eff}$ is the effective index change. From this we confirm the refractive index of the material and its modulation due to increased carrier concentration.

Using the transmission data, we can estimate the refractive index change in ITO due to the carrier concentration change in this dual ITO stack. By sweeping the length of the active device, the difference between the length of device region on the straight bus and on the waveguide result in different resonant frequency shift. The change in resonance frequency of the ring resonator is related to the change in the effective refractive index of propagating mode ($\Delta\lambda_{res} = \frac{\Delta n_{eff}}{n_g}\lambda$; $n_g$ is the group index). By measuring the resonance frequency of the ring, we calculate the effective index difference $\Delta n_{eff} = 0.002$. Given the mode overlap factor of $\Gamma = 1/100$ (numerically obtained), we calculate the change in refractive index of ITO $\Delta n_{ITO} = 0.2$, which is in good agreement with the numerical simulation and remarkably match the spectroscopic ellipsometry fitting parameter for an increased carrier concentration of $2.8\times10^{20}$ cm$^{-3}$.

After cutting the RING, we still observe a weaker electro-optic modulation (ER = 0.25 dB/μm) effect in the same compact device footprint. Here, the modulated signal can be directed to a neighboring waveguide and still switch off the signal in the main waveguide.

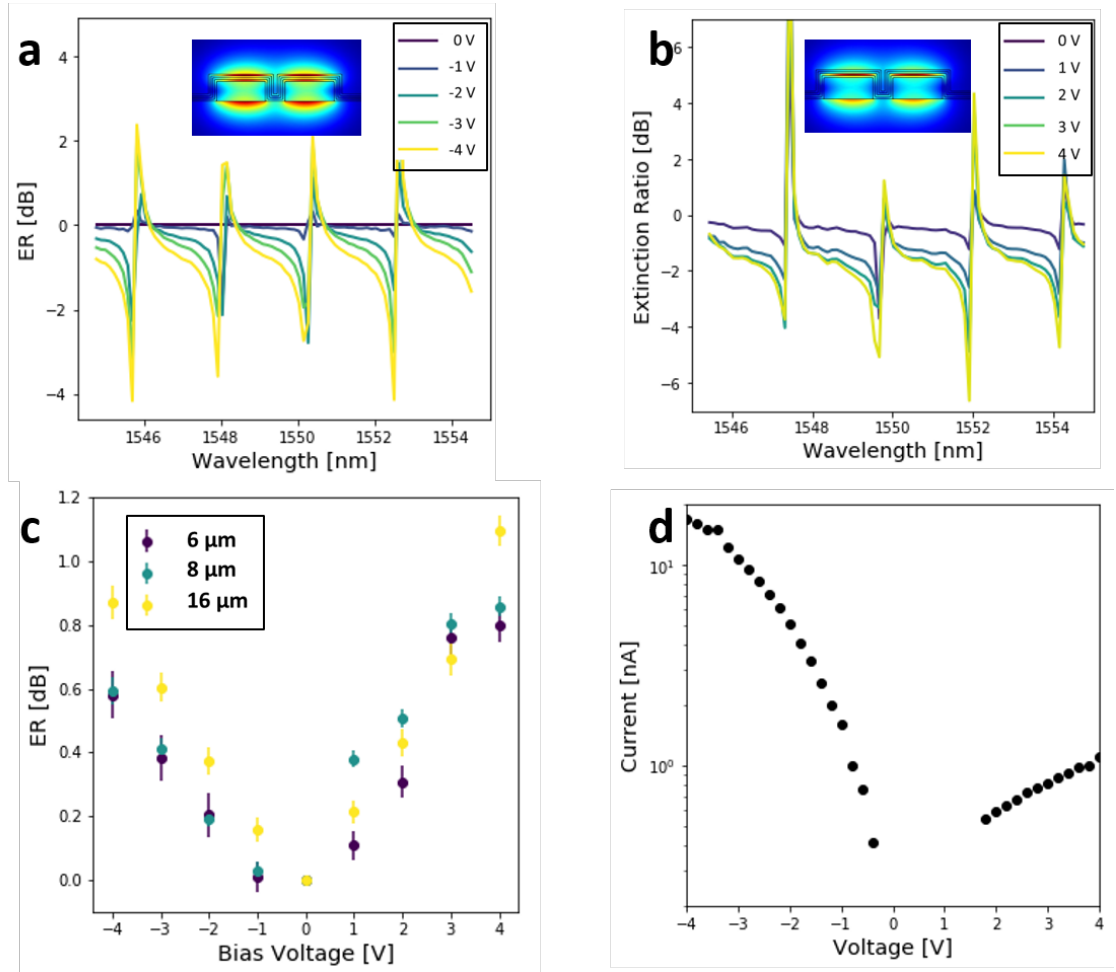

*Figure 4.* a, b, c Experimental measurements of the optical signal modulation performance in negative (a) and positive (b) bias, a TM mode grating coupler guides the incoming fiber coupled broadband EDFA signal into the waveguide. Output signal is then coupled out from another TM mode grating coupler into a tapered lens fiber and sent to an optical spectrum analyzer. The electrical bias is applied using a Source Measure Unit across to DC probe connected to the dual ITO layers. d) We use the same SMU unit to read the current flowing though the device to note IV characteristic and the breakdown voltage of the ALD $Al_2O_3$ gate oxide to be ~4.35 V. Measured transmission response under forward bias (inner ITO accumulation) show up to ~4 dB modulation depth for a 4μm long device. Modulation is observed at both polarities with a negative bias showing a smaller modulation depth due to smaller modal overlap at out ITO accumulation condition. By changing the coupling factor of the add-drop ring on one side and maintaining the critical coupling condition (a=r1) and taking into account the 0.13 nm shift in resonant wavelength due to change of refractive index in ITO under electric potential.

**CONCLUSIONS**

We experimentally demonstrate a coupling-controlled dual-gated ITO modulator heterogeneously integrated at the coupling regime between a silicon waveguide bus and a low-to-mid quality ($Q$)-factor microring resonator. We show a 4 dB extinction ratio (ER) modulation on resonance and 1.5 dB of modulation off resonance having an insertion loss of just 0.15 dB enabling a high ER/IL = 10:1 yet compact (4 μm) modulator. Careful process control enables an ITO material away from the ENZ-point to reduce losses yet demonstrate a decent ITO EO index change of 0.2. Taken together, ITO-silicon heterogeneous integration of an emerging EO material, ITO, which is commonly used in the semiconductor industry offers positive device-synergies demonstrating an compact coupling-based modulator on a silicon platform.

## METHODS

**RF Deposition:** ITO ultra-thin films were deposited on a cleaned Si substrate with a nominal 300 nm $SiO_2$ on it (1cm x 1cm) at 313K by reactive RF sputtering using Denton Vacuum Discovery 550 Sputtering System. They were prepared with the same time which is 1000 seconds. The target is consisting of 10% $SnO_2$ and 90% In2O3 by weight. The ITO films were prepared with the Argon and Oxygen flow-rates of 50 sccm. Deposition time is 1000s. There are four chips for each group. Two of them are tapped for later profilometer measurement. The vacuum setpoint is 5 Torr before deposition and the target will be pre-sputtered with the same deposition condition for 120s to remove the surface oxide layer of the target to avoid the contamination of the films. RF voltages were 300 voltage and RF bias were 25 voltage. After all parameters reached their set-points, deposition began.

**Ellipsometry:** We carried out spectroscopic ellipsometry measurement using J.A. Woollam M-2000 DI, which covered wavelength from 200 nm to 1680 nm. Analysis of the data used the corresponding CompleteEASE to extract thickness, complex optical constants, and other electrical parameters.

## ACKNOWLEDGEMENTS

VS is funded by AFOSR (FA9550-17-1-0377) and ARO (W911NF-16-2-0194) and HD by NASA STTR, Phase I (80NSSC18P2146).